\documentclass[prd,aps,superscriptaddress,floatfix,nofootinbib,eqsecnum]{revtex4-2}

\pdfoutput=1

\usepackage{amsmath}
\usepackage{amsfonts}
\usepackage{amsmath}
\usepackage{amssymb}
\usepackage{natbib}
\usepackage{bm}
\usepackage{adjustbox}
\usepackage{dcolumn}
\usepackage{graphicx}
\usepackage[utf8]{inputenc}
\usepackage{comment}
\usepackage{latexsym}
\usepackage{rotating}
\usepackage{hyperref}
\usepackage{subfigure}
\usepackage{color}
\usepackage[commandnameprefix=always]{changes}
\usepackage{verbatim}
\usepackage{comment}
\usepackage{empheq}
\usepackage{csquotes}
\usepackage{physics}
\usepackage{float}
\usepackage{soul}
\usepackage{amsmath,latexsym}
\usepackage{mathrsfs}
\usepackage{orcidlink}
\usepackage{booktabs}  
    \newcommand{\mykeywords}[1]{\textbf{Keywords:} #1}
    \newcommand{\pacscode}[2]{\textbf{PACS: }{ 98.80.Cq, 04.20.-q, 04.20.Cv} #2}
\begin{document}

\title{Inflation Driven by Scalar-Neutrino Coupling in a Mass-Varying Neutrino Framework}

\author{Hemanshi Bundeliya} \email {hemanshibundeliya07@gmail.com}\affiliation{Department of Physics,\\ Lovely Professional University, \\ Phagwara, Punjab, 144411, India}

\author{Gaurav Bhandari}\email{bhandarigaurav1408@gmail.com}\affiliation{Department of Physics,\\ Lovely Professional University, \\ Phagwara, Punjab, 144411, India}

\author{S. D. Pathak}\email{sdpathak@lko.amity.edu}\affiliation{Amity School of Applied Sciences,\\ Amity University Uttar Pradesh \\Lucknow Campus, Lucknow, 226228, India}

\author{V. K. Sharma}\email{vipin.33912@lpu.co.in}\affiliation{Department of Physics,\\ Lovely Professional University, \\ Phagwara, Punjab, 144411, India}

\begin{abstract}
We propose a cosmological framework in which neutrino masses evolve dynamically through coupling with a scalar field that simultaneously drives inflation. The neutrino mass is modeled as a power-law, exponential, or hybrid function of the scalar field, yielding an effective potential that includes neutrino backreaction. Starting from the Einstein–Hilbert action in a flat FLRW background, we derive the modified Friedmann and Klein–Gordon equations incorporating this coupling. Using the Fermi–Dirac integrals, we account for the continuous transition of neutrinos from relativistic to nonrelativistic regimes. The inflationary dynamics are analyzed via the slow-roll parameters derived from the effective potential. Our results show that the scalar–neutrino coupling alters the potential’s slope and curvature, thereby influencing the duration of inflation. The hybrid coupling form provides the most flexible realization, unifying neutrino mass generation with early-universe inflation within a single scalar-field framework.

\end{abstract}

\maketitle



\mykeywords{Scalar field Coupling, Mass-Varying Neutrino, Inflation}\\

\pacscode{98.80.Cq, 04.20.-q, 04.20.Cv}
\section{Introduction} \label{int}
Over the past quarter-century, a wide variety of cosmological observations has consolidated the standard concordance model of cosmology ($\Lambda$CDM): a spatially flat universe with an energy budget today composed of $\sim$ 5\% baryonic matter,~ $\sim$ 25\% cold dark matter (CDM), $\sim$ 70\% dark energy (DE) in the form of Einstein's positive cosmological constant ($\Lambda$) and smaller contributions provided by massive neutrinos and radiation~\cite{aghanim2020planck,DiValentino:2024xsv,Lesgourgues_2012,Hannestad_2010}. More recently, the baryon acoustic oscillation (BAO) dataset from the DE Spectroscopic Instrument (DESI) \cite{desi2025a,desi2025b} has provided new insights into the understanding of DE (see \cite{choudhury2024updated,Chaudhary:2025uzr,paliathanasis2025dark}).  Alongside these observational developments, theoretical studies have explored how modifications in the scalar-field Lagrangian and quantum gravity corrections can influence the dynamics of dark energy \cite{kaur2025dynamics,bhandari2025distortion}.

Another important avenue of investigation is the neutrino physics background. The hot Big Bang paradigm predicts a cosmic neutrino background: relic neutrinos that behave as radiation at early times and contribute as non-relativistic matter at late times \cite{Dolgov_2002,LESGOURGUES_2006,Weinberg:1962zza}. These relics affect the acoustic oscillations of the primordial plasma and the subsequent growth of large-scale structure, making cosmological observations sensitive to the total neutrino mass sum $\Sigma m_{\nu}$ and to the effective number of relativistic species $N_{\rm eff}$ \cite{lesgourgues2006}. Joint analyses that combine DESI with \textit{Planck} and supernova data currently tighten the upper bound on $\Sigma m_{\nu}$ to roughly $0.06\text{--}0.1\ \mathrm{eV}$ \cite{desi2025a,desi2025b}, while any significant deviation of $N_{\rm eff}$ from the standard value $3.044$ would indicate the presence of additional light relics \cite{grohs2016}.  In a recent work \cite{Chaudhary:2025uzr}, the authors
have obtained observational constraints on it.

Moreover,  the mechanism behind neutrino mass generation remains one of the most fundamental open questions in modern physics. Although the Standard Model originally treated neutrinos as massless, neutrino oscillation experiments have unambiguously shown that they possess small but non-zero masses \cite{PDG2024,Giganti_2018,Gonzalez_Garcia_2008}. Proposed explanations — including various seesaw realizations and radiative mass-generation mechanisms provide theoretical pathways for generating these masses \cite{brookfield2006cosmology,gellmann2013complexspinorsunifiedtheories,Yanagida:1980xy,PhysRevLett.44.912}, yet the fundamental origin remains an open problem. This issue acquires additional cosmological significance because scenarios such as time-dependent neutrino masses or novel neutrino couplings could leave observable imprints on background evolution and structure formation.

In this work, we propose a unified framework in which the neutrino mass is dynamically generated through coupling to a scalar field $\phi$, which simultaneously plays a role in the dynamics of the early universe. This idea is inspired by \textit{ mass-varying neutrinos} (MaVaN) scenarios, which have been studied primarily in the context of dark energy ~\cite{Fardon:2003eh, Gu:2003er}. However, their potential relevance during the inflationary epoch has not been extensively explored. This motivates our investigation into whether scalar-neutrino interactions can not only produce time-dependent neutrino masses but also influence or even drive cosmic inflation.

We construct a model where the neutrino mass is a function of the scalar field, specifically power law form ($ m_\nu(\phi) = m_0 \phi^n$); an exponential form ( $m_\nu(\phi) = m_0 e^{\beta \phi} $); 
or a hybrid form ($  m_\nu(\phi) = m_0 \phi^n e^{\beta \phi}$), where $m_0$ is the neutrino mass scale and $\beta$ represents the dimensionless coupling strength between the neutrino and the scalar field.
Such functional dependencies naturally arise in particle physics models featuring Yukawa couplings to scalar fields, as in quintessence or dilaton-type theories~\cite{Farrar:2003uw, amendola2000}. When neutrinos are coupled in this manner, they contribute additional terms to the energy-momentum tensor of the scalar field, modifying its dynamics through an effective potential (\cite{brookfield2006cosmology, Fardon:2003eh}):
\begin{equation}
    V_{\text{eff}}(\phi) = V(\phi) + m_\nu(\phi)(\rho_\nu - 3p_\nu),
\end{equation}
where $\rho_\nu$ and $p_\nu$ are the energy density and pressure of the neutrino fluid, respectively. Their evolution, especially during the transition from relativistic to nonrelativistic regimes, plays a crucial role in determining the behaviour of $\phi$.

We first examine whether this effective potential can sustain inflation by evaluating the slow-roll parameters $\epsilon(\phi)$ and $\eta(\phi)$, derived from $V_{\text{eff}}(\phi)$ with different $m_\nu(\phi)$. The analysis explores the conditions under which the effective potential remains sufficiently flat and the scalar field evolves slowly, as required for successful inflation.

In a complementary and reverse approach, we start from a known inflationary potential (e.g., $V(\phi) \propto \phi^2$, plateau models such as Starobinski inflation~\cite{Starobinsky:1980te}, or $\alpha$-attractors~\cite{Kallosh_2013}) and attempt to reconstruct the corresponding functional form of $m_\nu(\phi)$ that would yield the same effective potential. This ``inverse method'' provides a novel way to link inflationary dynamics with neutrino mass generation mechanisms.

By combining both forward and reverse analyzes, we investigated the theoretical consistency and phenomenological viability of a model in which mass-varying neutrinos actively participate in the dynamics of the early universe. This approach opens an intriguing possibility: that a single scalar field might govern multiple cosmological epochs, driving inflation, modulating neutrino mass, and potentially contributing to late-time cosmic acceleration. As such, this framework provides a natural bridge between the physics of the early universe, the structure of the neutrino sector, and dark-energy models.

The paper is organized as follows. In Sect.\ref{bm}, we discussed the theoretical framework and the derivation of the effective potential for various forms of neutrino mass coupled with a quintessence scalar field. In Sect.\ref{s1}, we analyze the inflationary scenarios using slow-roll parameters based on the effective potential. Finally, we summarize our findings in Sec. \ref{con}.

We prefer to work with (+ - - -) metric signature, and the natural units $c=\hbar=1$. 
 
\section{Background mathematical formalism and couplings}\label{bm}
We consider the $4-$dimensional Einstein-Hilbert action  as
\begin{eqnarray} \mathcal{S}= \frac{1}{2\kappa^{2}}\int d^{4}x \sqrt{-g} \ R+ \mathcal{S}_{m}(g_{\mu\nu}, \Psi_{m})\label{a1}\end{eqnarray}
where $\kappa^{2}=8\pi G$ and $\mathcal{S}_{m}$ is the action of the matter part with matter field $\Psi_{m}$.
We assume the spacetime as homogeneous, isotropic and spatially flat. It is given by the Friedmann-Lemaitre-Robertson-Walker (FLRW) spacetime as
\begin{eqnarray}ds^{2}= -dt^{2} + a^2(t)[dr^2 + r^2 (d\theta^{2} + \sin^{2}\theta d\phi^{2})]\label{a242} \end{eqnarray}
where $a(t)$ is time dependent scale factor and the speed of light $c=1$.

  Now, varying the action \eqref{a1} with respect to the metric tensor $g^{\mu\nu}$ yields the famous Einstein field equation   
\begin{equation}\label{field}
    R_{\mu\nu} -\frac{1}{2}Rg_{\mu\nu}=8\pi G T_{\mu\nu},
\end{equation}
where  $R_{\mu\nu}$ is the Ricci tensor, $R$  is the Ricci scalar and $T_{\mu \nu }$ represents the stress-energy tensor fields defined as 
\begin{align}
   T_{\mu\nu} &\equiv-\frac{2}{\sqrt{-g}}\frac{\delta (\sqrt{-g} \mathcal{L})}{\delta g^{\mu\nu}}  \label{stressm},
   \end{align} 




The conservation of local energy momentum tensor gives the continuity equation as \begin{equation}\label{n6}
\dot{\rho}_{\text{total}}+3H(1 + w)\rho_{\text{total}}=0.
\end{equation}

The action \eqref{a1} can be recast as a scalar field minimally coupled to General Relativity as
\begin{equation}\label{actionneu}
S = \int d^4x \, \sqrt{-g} 
\left[ 
\frac{1}{2} M_P^2 R 
- \frac{1}{2} \partial_\mu \phi \, \partial^\mu \phi 
- V(\phi)
\right] 
+ \sum_j S_j \big[ B_j^2(\phi) g_{\mu\nu}, \psi_j \big],
\end{equation}
where $\phi(t)$ represents a homogeneous scalar field (quintessence) with potential $V(\phi)$, and $\psi_j$ denotes other matter fields. Here $M_P = 2.4 \times 10^{18}\,\mathrm{GeV}$ is the reduced Planck mass.
The coupling between the scalar field and matter fields is introduced through a conformal transformation of the metric, 
$B_j^2(\phi) g_{\mu\nu}$, where $B_j(\phi) > 0$. 

The first Friedmann equation takes the form
\begin{equation}
H^2 = \frac{1}{3M_P^2} 
\left( 
\frac{1}{2}\dot{\phi}^2 + V + \sum_j \rho^{(j)} 
\right),
\label{eq:friedmann}
\end{equation}
which relates the Hubble parameter $H = \dot{a}/a$ to the total energy density of the scalar field and all other matter components.
The time evolution of the Hubble parameter is given by
\begin{equation}
\dot{H} = -\frac{1}{2M_P^2}
\left(
\dot{\phi}^2 + \sum_j (\rho^{(j)} + P^{(j)})
\right),
\label{eq:hubdot}
\end{equation}
indicating that the kinetic term of the scalar field, together with the pressure of different species, governs the deceleration of the Universe.

The conservation equations for each component of the Universe are modified in the presence of an interaction between the scalar field and matter. 
For an interacting species $i$, the continuity equation takes the form
\begin{equation}
\dot{\rho}^{(i)} + 3H\big(\rho^{(i)} + P^{(i)}\big) 
= \frac{B_{,\phi}}{B} \, \dot{\phi} \, \big(\rho^{(i)} - 3P^{(i)}\big),
\label{eq:conti_int}
\end{equation}
indicating an exchange of energy between the scalar field and the coupled matter sector through the conformal function $B(\phi)$. 
For other noninteracting components $j$, the standard conservation equation holds:
\begin{equation}
\dot{\rho}^{(j)} + 3H\big(\rho^{(j)} + P^{(j)}\big) = 0,
\label{eq:conti_nonint}
\end{equation}
which represents the usual adiabatic dilution of energy density with the expansion of the Universe.The interaction between neutrinos and the quintessence field obtained from Eq.(\ref{actionneu}) can also be interpreted within the framework of the coupled quintessence models 
discussed in Refs.~\cite{amendola2000, Farrar_2004,wetterich1994cosmonmodelasymptoticallyvanishing}.

\subsection{Continuity equation for Mass-Varying Neutrino}

The neutrino mass is assumed to depend on the scalar field, 
$m_\nu = m_\nu(\phi)$. 
In cosmology, neutrinos must be described kinetically rather than as a perfect fluid. 
Their dynamics are governed by the phase-space distribution function 
$f(x^i, p^i, t)$, 
which satisfies the collisionless Boltzmann (Liouville) equation in an expanding universe \cite{Brookfield_2006}
\begin{equation}\label{botz}
\frac{df}{dt} = 
\frac{\partial f}{\partial t} 
+ \dot{x}^i \frac{\partial f}{\partial x^i} 
+ \dot{p}^i \frac{\partial f}{\partial p^i} = 0.
\end{equation}

For a homogeneous and isotropic background, the distribution function $f$ depends only on the magnitude of the comoving momentum, 
$q = a p$. 
Solving this equation gives the following expressions for the neutrino energy density and pressure:
\begin{equation}\label{neurho}
\rho_\nu = \frac{1}{a^4} \int q^2 \, dq \, d\Omega \, \epsilon(q) \, f_0(q),
\qquad
p_\nu = \frac{1}{3a^4} \int q^2 \, dq \, d\Omega \, f_0(q) \, \frac{q^2}{\epsilon(q)},
\end{equation}
The above expressions represent the general definitions of the neutrino energy density and pressure in terms of the phase-space distribution function. Here, the comoving momentum $q=ap$ remains constant in the absence of interactions, and the energy–momentum relation explicitly incorporates the mass variation through $m_{\nu}(\phi)$. 

where $f_0(q)$ is the background Fermi--Dirac distribution function, and the single-particle energy satisfies 
$\epsilon^2 = q^2 + a^2 m_\nu^2(\phi)$. 
Now, assuming isotropy $(\int d\Omega = 4\pi)$ and introducing suitable dimensionless variables, we can evaluate the integrals for the neutrino energy density and pressure in a compact form
\begin{equation}
z = \frac{\epsilon}{T_\nu}, 
\qquad 
\zeta = \frac{m_\nu(\phi)}{T_\nu},
\end{equation}
with the neutrino temperature scaling as $T_\nu(a) = T_{\nu,0}/a$, 
The energy density and pressure of neutrinos can be expressed as
\begin{equation}
\rho_\nu(a) = \frac{2}{\pi^2} \, T_\nu^4(a) \, I_\varepsilon(\zeta),
\qquad
p_\nu(a) = \frac{2}{3\pi^2} \, T_\nu^4(a) \, I_{3/2}(\zeta),
\end{equation}
Here, $I_\varepsilon(\zeta)$ and $I_{3/2}(\zeta)$ denote the standard Fermi–Dirac integrals, whose forms encapsulate the transition between the relativistic and nonrelativistic regimes of the neutrino population.

This integrals are defined as
\begin{equation}
I_\varepsilon(\zeta) = 
\int_{\zeta}^{\infty} 
\frac{z^2 \sqrt{z^2 - \zeta^2}}{e^z + 1} \, dz,
\qquad
I_{3/2}(\zeta) = 
\int_{\zeta}^{\infty} 
\frac{(z^2 - \zeta^2)^{3/2}}{e^z + 1} \, dz.
\end{equation}
The above integrals do not admit closed-form expressions for arbitrary $\zeta$ but their limiting forms can be obtained in the relativistic$(\zeta << 1)$ and nonrelativistic $(\zeta \gtrsim 1)$ limits.

Specifically, one finds for $I_{3/2}(\zeta)$,corresponding to the pressure integral, the asymptotic expansions read,
\begin{equation}
I_{3/2}(\zeta) =
\begin{cases}
\dfrac{7\pi^4}{120} - \dfrac{\pi^2}{8}\zeta^2 + \mathcal{O}(\zeta^4 \log \zeta), & \zeta < 1, \\[8pt]
3\zeta^2 K_2(\zeta) + \mathcal{O}(e^{-2\zeta}), & \zeta \gtrsim 1.
\end{cases}
\end{equation}
where $K_\nu(x)$ is the modified Bessel function of second kind. Similarly, one finds,
\begin{equation}
I_\varepsilon(\zeta) =
\begin{cases}
\dfrac{7\pi^4}{120} - \dfrac{\pi^2}{24}\zeta^2 + \mathcal{O}(\zeta^4 \log \zeta), & \zeta < 1, \\[8pt]
3\zeta^2 K_2(\zeta) + \zeta^3K_1(\zeta)+\mathcal{O}(e^{-2\zeta}), & \zeta \gtrsim 1.
\end{cases}
\end{equation}
One easily finds that for the relativistic limit $(\zeta \ll 1)$, these integrals reproduce the standard results,
\begin{equation}
\rho_\nu \simeq N_F \frac{7\pi^2}{60} T_\nu^4,
\qquad
p_\nu = \frac{\rho_\nu}{3}.
\end{equation}
These relations ensure consistency with the radiation-dominated behavior at early times, while deviations at large $(\zeta)$ encode the impact of neutrino mass generation and its coupling to the scalar field.

From the expression for energy density and pressure in Eq.\eqref{neurho}, the evolution of energy density of neutrinos is derived as
\begin{equation}
\dot{\rho}_\nu + 3H(\rho_\nu + p_\nu) = \frac{\partial \ln{m_\nu(\phi)}}{\partial \phi} \dot{\phi} (\rho_\nu - 3p_\nu).
\end{equation}
The above equation describes the non-conservation of the neutrino energy density due to its explicit dependence on the scalar field. 
The interaction term on the right-hand side quantifies the energy transfer between the neutrino fluid and the scalar field through the field-dependent neutrino mass. 
A positive coupling implies that the scalar field loses energy to the neutrino sector as the field evolves.

The total energy of the combined system, however, must remain conserved. 
Hence, when both the scalar field and neutrino components are considered together, their total energy density satisfies the conservation equation
\begin{equation}
\sum_i \dot{\rho}_i + 3 H \sum_i (\rho_i + p_i) = 0,
\end{equation}
which ensures that the total energy–momentum tensor of the system is covariantly conserved.

From this total conservation law, one obtains the modified Klein–Gordon equation governing the evolution of the scalar field in the presence of the coupling to neutrinos:
\begin{equation}\label{massvarklein}
\ddot{\phi} + 3H\dot{\phi} +  \frac{\partial V}{\partial \phi} = -  \frac{\partial \ln{m_\nu}}{\partial \phi} (\rho_\nu - 3p_\nu)
\end{equation}
From this equation, the dynamics of the field is specified by the effective potential.

\begin{equation}
V_{\text{eff},\phi} = V_{,\phi} + (\tilde{\rho}_\nu - 3\tilde{p}_\nu)m_{\nu,{\phi}}(\phi)
\end{equation}
where $\tilde{\rho}_\nu=\rho_\nu/m_\nu(\phi)$ and $\tilde{p}_\nu= p_\nu/m_\nu(\phi)$. 


\begin{equation}\label{eq:12}
V_{\text{eff},\phi} = V_{,\phi} + \frac{m_{\nu,\phi}(\phi)}{m_{\nu}(\phi)} \left( \rho_{\nu} - 3p_{\nu} \right).
\end{equation}

Using the total energy conservation equation for the combined fluids, 
the Friedmann and Raychaudhuri equations take the form

\begin{eqnarray}
&&H^2 = \frac{8\pi G}{3} \sum_i \rho_i, \label{fridtotal}\\
&&\frac{\ddot{a}}{a} = -\frac{4\pi G}{3} \sum_i \left( \rho_i + 3p_i \right). \label{raytotal}
\end{eqnarray}

Where, \begin{eqnarray}
\sum_i \rho_i &=& \rho_\nu + \rho_\phi, \\
\sum_i (\rho_i + 3 p_i) &=& \rho_\nu + \rho_\phi + 3 p_\nu + 3 p_\phi
\end{eqnarray}


The above framework establishes the background dynamics of a cosmological model where neutrinos interact with a scalar field through a mass-varying coupling. 
The coupling modifies the evolution of both components by introducing an additional source term in the scalar field equation and redefining its potential energy through the effective potential $V_{\text{eff}}(\phi)$. 
This effective potential governs the field’s dynamics and can significantly influence the early- and late-time evolution of the Universe. 
In particular, its form determines whether the scalar field can drive an inflationary phase or act as a dark energy component at late times. 
In the following section, we analyze the structure of this effective potential and explore its implications for inflationary dynamics within the mass-varying neutrino framework.

\section{Evolution of slow roll parameters with effective potential}\label{s1}
In this subsection, we consider different functional forms of the neutrino mass and derive the corresponding expressions for the slow-roll parameters and their evolution\cite{Liddle:2000cg,Guth:1981,Planck:2018jri}. 
To achieve a phase of sudden accelerated expansion of about 60 e-folds within the framework of the Raychaudhuri equation, the first slow-roll condition is imposed, which ensures that the potential energy of the scalar field dominates over its kinetic energy as
\begin{equation}\label{3seca}\frac{1}{2}\dot{\phi}^2 \ll V(\phi) \end{equation}
This inequality ensures that the energy density is large enough and nearly constant to drive inflation. Furthermore, to ensure that inflation persists for an extended period, the dynamics of the scalar field must be governed predominantly by the Hubble friction term
\begin{equation}|\ddot{\phi}|\ll |2H\dot{\phi}|.\end{equation}
These conditions collectively ensure that the inflaton rolls slowly along its potential, giving rise to an approximately de Sitter phase characterized by a quasi-constant Hubble rate. 

From Eq.(\ref{3seca}) and Eq.(\ref{massvarklein}), the slow roll condition is satisfied when 
\begin{equation}
    \frac{1}{2}\left[\frac{V_{eff,\phi}}{3H\{1+\ddot{\phi}/3H\dot \phi)\}}\right]^2\ll V(\phi),
\end{equation}
which gives the first slow roll parameter, denoted by $\epsilon'$, defined as 
\[
\epsilon' = \frac{1}{2}\left( \frac{V'_{eff}(\phi) }{V(\phi)} \right)^2,
\] 
where $V(\phi)$ is the inflaton potential and $V'(\phi)$ is its derivative with respect to the field $\phi$. A small value of $\epsilon'$ (that is, $\epsilon' \ll 1$) indicates that the Hubble parameter is nearly constant, which is a key requirement to sustain inflation. Similarly, the second slow-roll parameter, $\eta$, measures the influence of the acceleration of the inflaton on the inflationary dynamics. It is defined as
\begin{equation}
    \eta = \frac{V_{eff}''(\phi)}{V(\phi)},
\end{equation}

where $V''(\phi)$ is the second derivative of the potential. Inflation continues as long as $\eta \ll 1$, ensuring that the inflaton rolls slowly enough for the universe to expand exponentially. Collectively, these parameters characterize the flatness and shape of the inflaton potential and play a crucial role in predicting observable quantities such as the scalar spectral index and tensor-to-scalar ratio.

To get more intuition of the model we consider a Gaussian-type potential \cite{Sadjadi_2017} and different mass terms of neutrinos as
\begin{eqnarray}
    V(\phi) &=& V_0 (1 - e^{-\alpha \phi^2}), \nonumber \\
    m_\nu(\phi) &=&
    \left\{
    \begin{array}{l}
        m_0 e^{\beta \phi}, \\
        m_0 \phi^n, \\
        m_0 \phi^n e^{\beta \phi}.
    \end{array}
    \right.
\end{eqnarray}
where $\alpha$ and $\beta$ are the positive constants and $m_0$ is the neutrino mass at $\phi=0$.

During the inflationary epoch, it is common to approximate $\rho_\nu - 3p_\nu \approx 0$, assuming that neutrinos remain ultrarelativistic and their contribution to the effective potential is negligible. However, in the present analysis, we relax this approximation and explicitly retain the trace term by solving the Fermi–Dirac integrals for the neutrino energy density and pressure in the relativistic-to-nonrelativistic transition regime. Using the approximate relations
\begin{align}
\rho_\nu &\simeq \frac{1}{\pi^2 \beta^4}\left(\frac{7\pi^4}{120} - \frac{\pi^2}{24}\left(\frac{m_\nu(\phi)}{T_\nu}\right)^2\right), \\[4pt]
p_\nu &\simeq \frac{1}{3\pi^2 \beta^4}\left(\frac{7\pi^4}{120} - \frac{\pi^2}{8}\left(\frac{m_\nu(\phi)}{T_\nu}\right)^2\right),
\end{align}
the trace contribution is obtained as
\begin{equation}
\rho_\nu - 3p_\nu \simeq \frac{1}{12}\left(\frac{m_\nu(\phi)}{T_\nu}\right)^2.
\end{equation}
In the following, we perform a case-by-case analysis corresponding to different functional dependencies of the neutrino masses on the scalar field. 

\subsection{Case I: Exponential Potential, $m_\nu(\phi)={m_0e^{\beta \phi}}$}
\begin{figure}[H]
    \centering
    \includegraphics[width=0.6\textwidth, height=0.3\textheight]{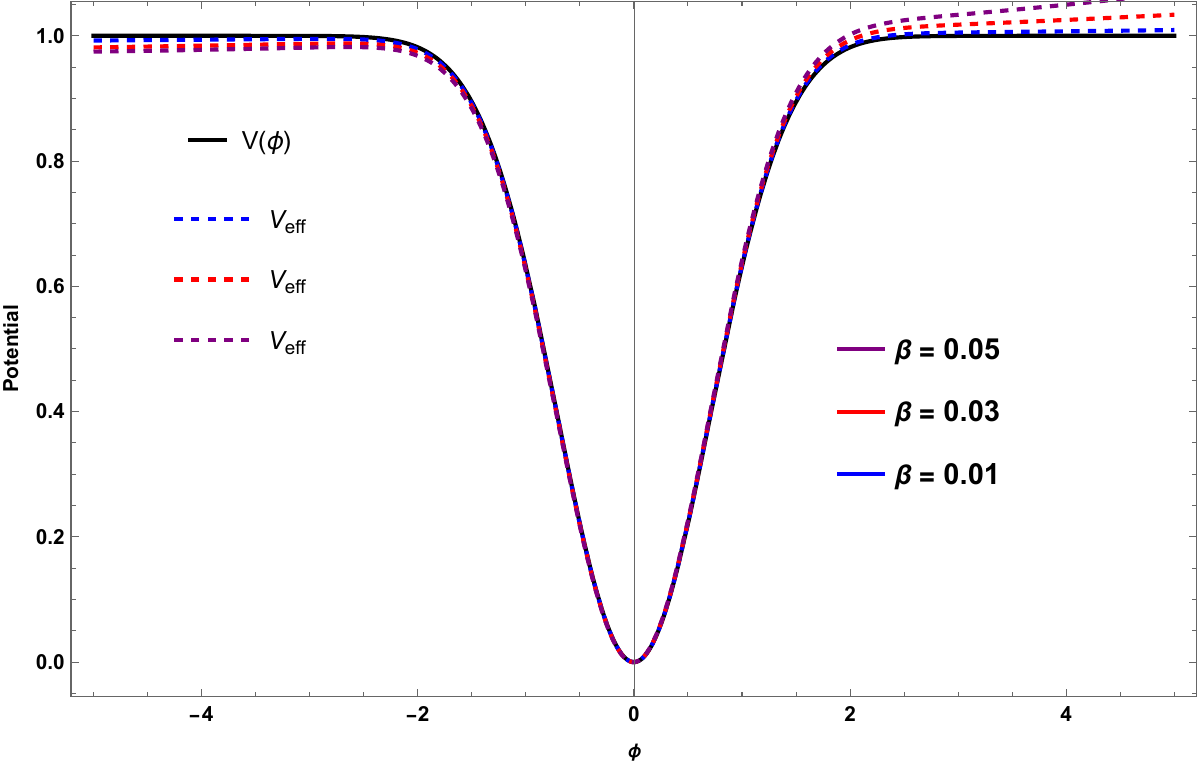}
    \caption{Comparison of $V(\phi)$ and $V_{\text{eff}}(\phi)$ with $m_\nu=m_0e^{\beta \phi}$ for different values of the coupling parameter $\beta$, with $V_0=1$ and $\alpha=1$.}
    \label{exponential}
\end{figure}

We know from Eq.~(\ref{eq:12}) that the scalar potential acquires a correction due to its coupling with the neutrino mass, leading to a modified effective potential. Figure~\ref{exponential} illustrates the behavior of the scalar field potential $V(\phi)$ and the corresponding effective potential $V_{\text{eff}}(\phi)$ for the exponential mass coupling case, $m_\nu(\phi) = m_0 e^{\beta \phi}$. As depicted in the figure, the effective potential progressively deviates from the original potential as the coupling parameter $\beta$ increases. For small values of $\beta$, $V_{\text{eff}}(\phi)$ closely traces $V(\phi)$, reflecting a weak interaction between the scalar field and the neutrino sector. However, as $\beta$ grows, the coupling contribution $\beta(\rho_\nu - 3p_\nu)$ becomes increasingly significant, leading to a steeper and slightly elevated effective potential at larger field amplitudes. This behavior indicates that a stronger scalar–neutrino coupling enhances both the slope and curvature of the potential, thereby raising the effective energy scale and altering the slow-roll dynamics. Consequently, larger values of $\beta$ tend to make the slow-roll conditions more difficult to satisfy, emphasizing the pivotal role of the coupling strength in shaping the inflationary evolution and the overall structure of the effective potential.

The slow-roll parameters are evaluated as,
\begin{equation}
\epsilon' = \frac{1}{2} 
\left(
\frac{
2V_0 \alpha \phi e^{-\alpha \phi^2}
+ \beta  \frac{1}{12}\left(\frac{m_\nu(\phi)}{T_\nu}\right)^2 
}{
V_0 \left( 1 - e^{-\alpha \phi^2} \right)
}
\right)^2.
\end{equation}

\begin{equation}
\eta =
\frac{
2V_0 \alpha e^{-\alpha \phi^2}(1 - 2\alpha \phi^2)
+ \beta^2 \frac{1}{12}\left(\frac{m_\nu(\phi)}{T_\nu}\right)^2 
}{
V_0 \left( 1 - e^{-\alpha \phi^2} \right)
}.
\end{equation}

\noindent
The slow-roll parameters $\epsilon'$ and $\eta$ are influenced by the coupling parameter $\beta$, which governs the effect of neutrino backreaction on the inflationary dynamics. For smaller $\beta$, the slow-roll conditions $\epsilon' < 1$ and $|\eta| < 1$ remain satisfied, ensuring a prolonged inflationary phase. As $\beta$ increases, both parameters approach unity more rapidly, leading to a faster end of inflation and a shorter slow-roll duration.

\begin{figure}[H]
\centering
\begin{minipage}[t]{0.45\textwidth}
\centering
\includegraphics[width=1.10\textwidth]{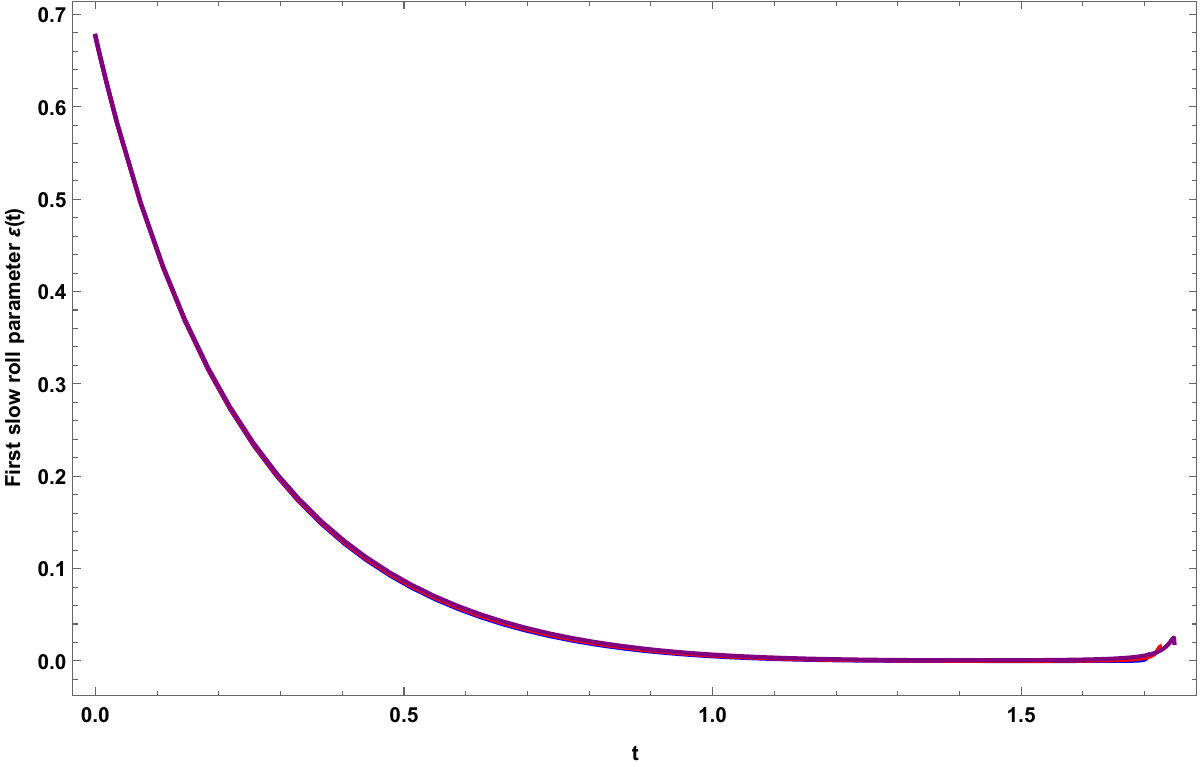}
\end{minipage}
\hfill
\begin{minipage}[b]{0.45\textwidth}
\centering
\includegraphics[width=1.10\textwidth]{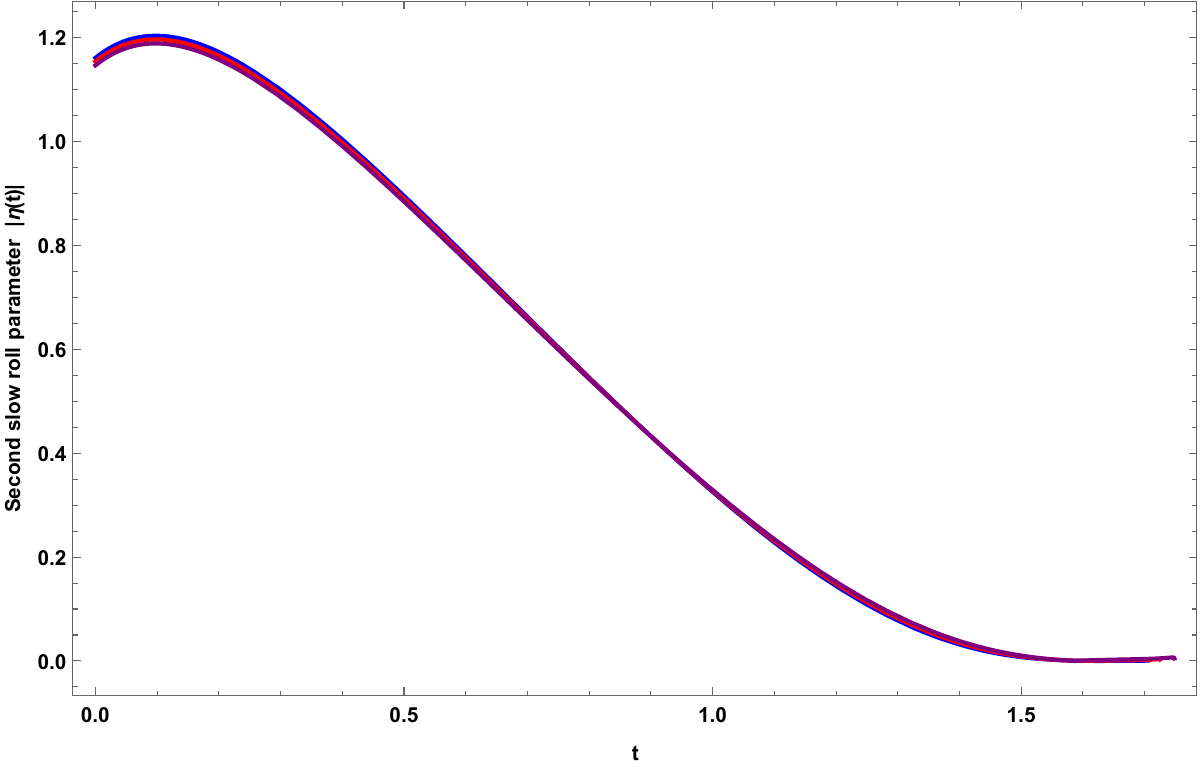}
\end{minipage}
\caption{Evolution of the first (\textit{left panel}) and second (\textit{right panel}) slow-roll parameters, $\epsilon'$ and $\eta$, respectively, for $V_0 = 1$ and $\alpha = 1$. The \textit{blue}, \textit{red}, and \textit{purple} curves correspond to coupling parameters $\beta = 0.01$, $\beta = 0.03$, and $\beta = 0.05$, respectively.}\label{slowexp}
\end{figure}

\noindent
Figure~\ref{slowexp} shows the evolution of the first and second slow-roll parameters, $\epsilon'$ and $\eta$, for the exponential coupling case $m_\nu(\phi) = m_0 e^{\beta \phi}$. The left panel presents the variation of the first slow-roll parameter, $\epsilon'$, with time for different values of the coupling parameter $\beta$. For smaller coupling values, such as $\beta = 0.01$, $\epsilon'$ remains nearly constant and small, indicating that the field stays in the slow-roll regime for a longer duration. As the coupling strength increases to $\beta = 0.03$ and $\beta = 0.05$, $\epsilon'$ rises more rapidly, showing that the scalar field experiences a steeper effective potential and evolves faster, which leads to an earlier end of inflation. The right panel illustrates the time evolution of the second slow-roll parameter, $\eta$, for the same set of coupling values. For small $\beta$, $|\eta|$ remains below unity, confirming the validity of the slow-roll approximation. However, as $\beta$ increases, $|\eta|$ grows faster and reaches higher values, reflecting an increase in the curvature of the effective potential due to stronger coupling. Overall, the figure demonstrates that increasing the coupling parameter $\beta$ enhances both the slope and curvature of the potential, making the scalar field roll faster and reducing the duration of the inflationary phase.

\subsection{Case II: Power Law, $m_\nu(\phi)={m_0 \phi^n}$}

\begin{figure}[H]
    \centering
    \includegraphics[width=0.6\textwidth, height=0.3\textheight]{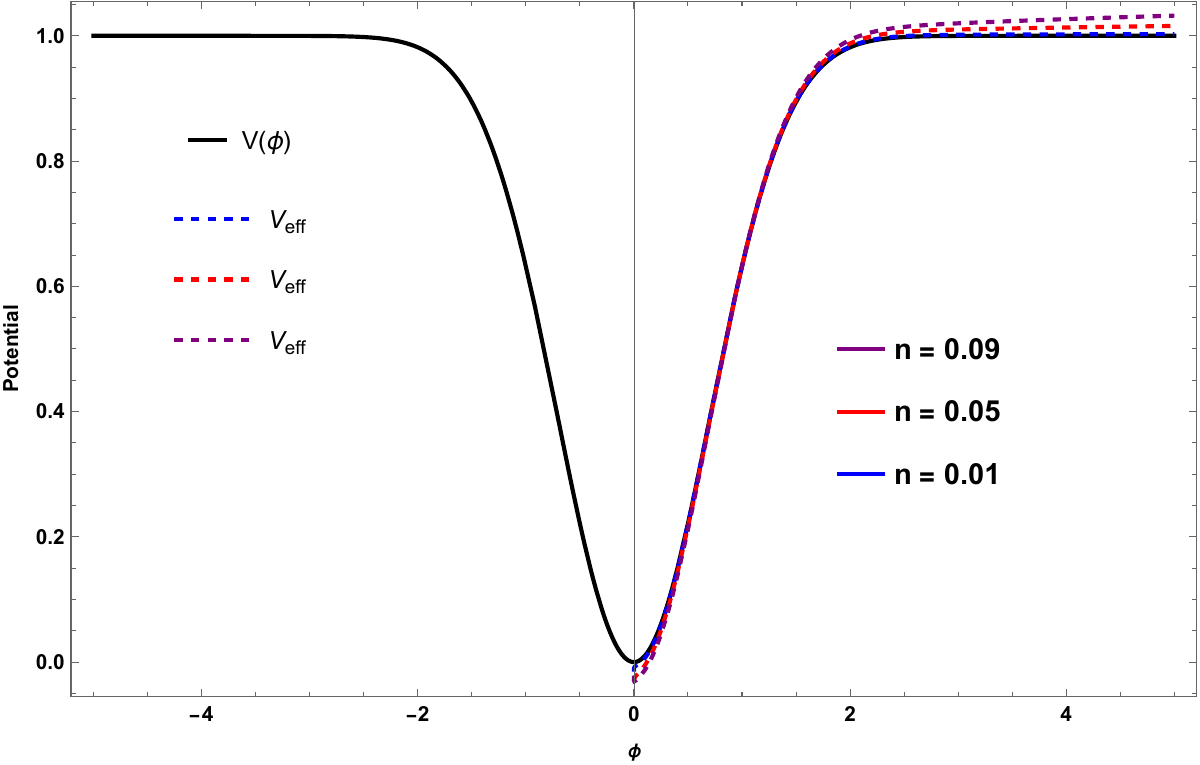}
    \caption{Comparison of $V(\phi)$ and $V_{\text{eff}}(\phi)$ with $m_\nu=m_0 \phi^n$ for different values of the coupling parameter $n$, with $V_0=1$ and $\alpha=1$.}
    \label{powerpotential}
\end{figure}

\noindent
Figure~\ref{powerpotential} presents the behaviour of the scalar field potential $V(\phi)$ and the effective potential $V_{\text{eff}}(\phi)$ for the power-law coupling case $m_\nu(\phi)=m_0 \phi^n$. In contrast to the exponential case, the deviation between $V(\phi)$ and $V_{\text{eff}}(\phi)$ increases more gradually with the coupling index $n$. For small $n$, the two potentials nearly coincide, implying minimal influence from the neutrino backreaction. As $n$ increases, the effective potential slowly rises above the original potential, showing a mild enhancement in slope at higher field values. This suggests that the power-law interaction modifies the potential in a smoother and more controlled manner, producing a slower departure from the standard inflationary trajectory. Hence, the coupling index $n$ acts as a gentle regulator of the field dynamics, influencing the duration of inflation without introducing strong distortions in the potential profile.

The slow-roll parameters are evaluated as,

\begin{equation}
\epsilon' = \frac{1}{2}
\left(
\frac{
2V_0 \alpha \phi e^{-\alpha \phi^2}
+ \frac{n}{\phi} \frac{1}{12}\left(\frac{m_\nu(\phi)}{T_\nu}\right)^2 
}{
V_0 \left( 1 - e^{-\alpha \phi^2} \right)
}
\right)^2.
\end{equation}

\begin{equation}
\eta =
\frac{
2V_0 \alpha e^{-\alpha \phi^2}(1 - 2\alpha \phi^2)
+ \frac{n(n - 1)}{\phi^2}
\frac{1}{12}\left(\frac{m_\nu(\phi)}{T_\nu}\right)^2 
}{
V_0 \left( 1 - e^{-\alpha \phi^2} \right)
}.
\end{equation}

\noindent
The slow-roll parameters $\epsilon'$ and $\eta$ are modified by the power-law coupling parameter $n$, which determines the strength of the neutrino--scalar interaction. A larger value of $n$ enhances the coupling, increasing both the slope and curvature of the effective potential. As a result, the slow-roll conditions are affected, leading to a faster evolution of the scalar field and an earlier end of inflation. This behaviour produces a distinct inflationary dynamics compared to the exponential coupling scenario.

\begin{figure}[H]
\centering
\begin{minipage}[t]{0.45\textwidth}
\centering
\includegraphics[width=1.10\textwidth]{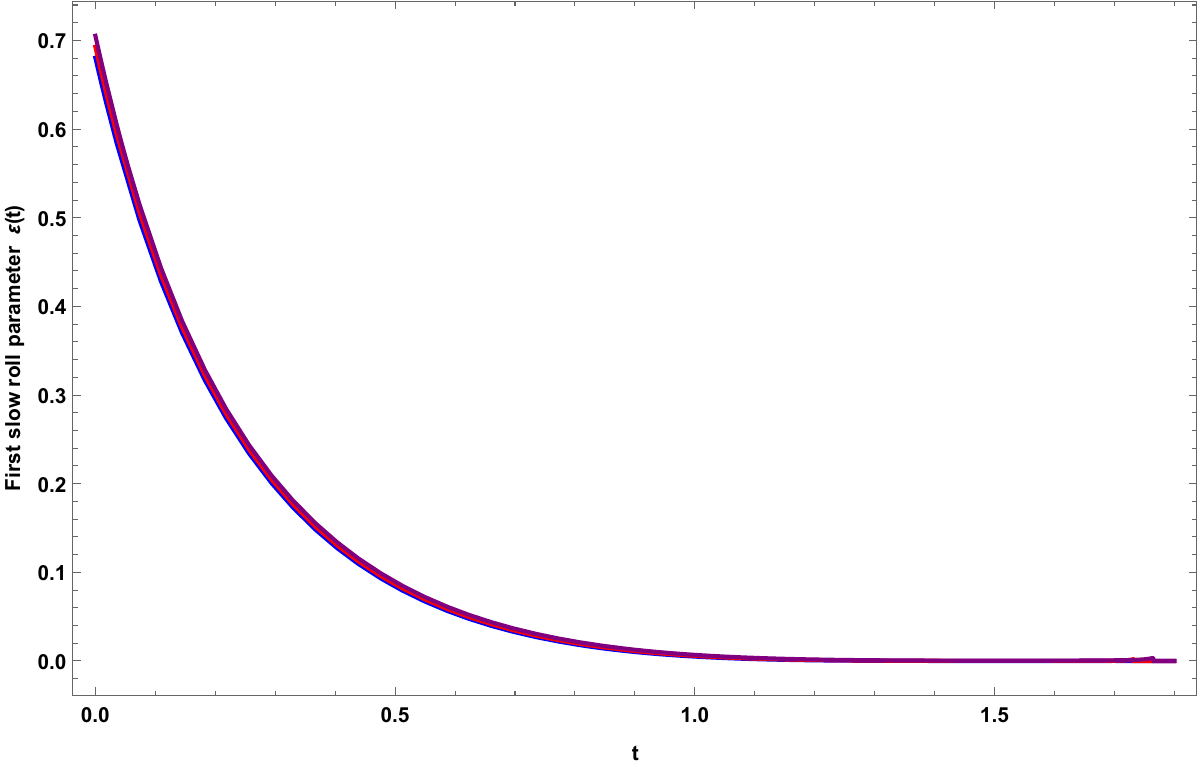}
\end{minipage}
\hfill
\begin{minipage}[b]{0.45\textwidth}
\centering
\includegraphics[width=1.10\textwidth]{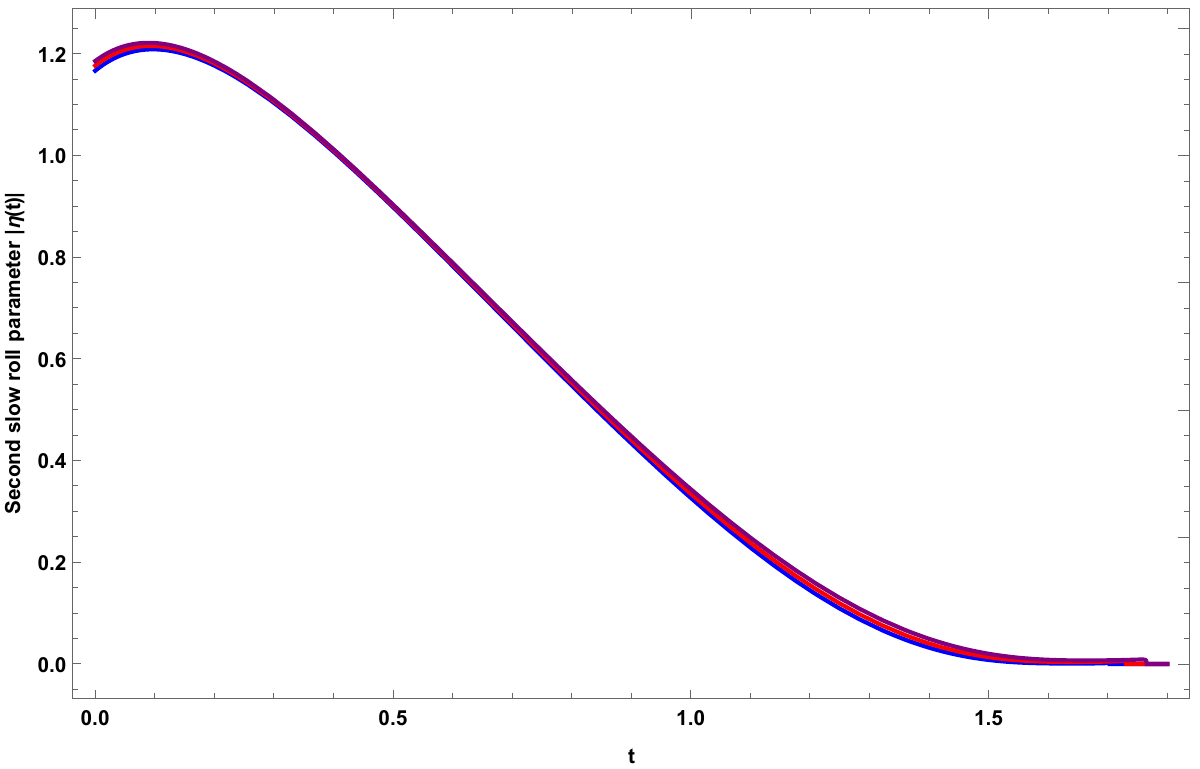}
\end{minipage}
\caption{Evolution of the first (left panel) and second (right panel) slow-roll parameters, $\epsilon'$ and $\eta$, respectively, for $V_0 = 0.1$ and $\alpha = 1$. The \textit{blue}, \textit{red}, and \textit{purple} curves correspond to the coupling parameters $n = 0.01$, $n = 0.03$, and $n = 0.05$, respectively.}\label{slowpower}
\end{figure}

\noindent
Figure~\ref{slowpower} displays the evolution of $\epsilon'$ and $\eta$ for the power-law coupling case $m_\nu(\phi) = m_0 \phi^n$. The qualitative behaviour of both slow-roll parameters is similar to the exponential case; however, the influence of the coupling index $n$ is more moderate. For small $n$, the field remains in a sustained slow-roll regime, while larger $n$ values increase both $\epsilon'$ and $\eta$, leading to a faster end of inflation. Compared to the exponential coupling, the variation in $\epsilon'$ and $\eta$ is smoother, suggesting that the power-law coupling produces a relatively gradual transition from slow-roll to the end of inflation. Thus, $n$ controls the steepness of the potential more gently than $\beta$, allowing for a slightly longer inflationary phase for comparable coupling strengths.
\subsection{Case III: Mixed Form, $m_\nu(\phi)=m_0 \phi^n e^{\beta \phi}$}

\begin{figure}[H]
    \centering
    \includegraphics[width=0.6\textwidth, height=0.3\textheight]{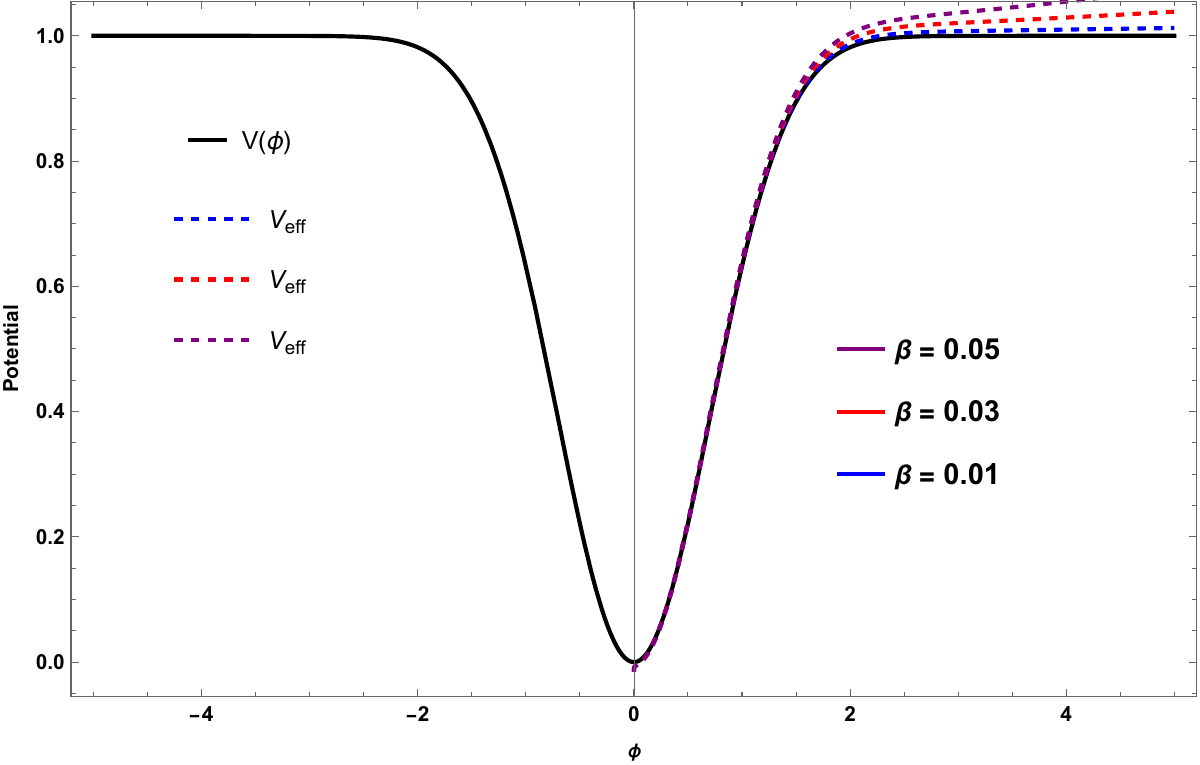}
    \caption{Comparison of $V(\phi)$ and $V_{\text{eff}}(\phi)$ with $m_\nu=\phi^n e^{\beta \phi}$ for different values of the coupling parameter $\beta$ and $n=0.01$, with $V_0=1$ and $\alpha=1$.}
    \label{combpotential}
\end{figure}

Figure~\ref{combpotential} illustrates the behaviour of the scalar field potential $V(\phi)$ and the effective potential $V_{\text{eff}}(\phi)$ for the mixed coupling case $m_\nu(\phi)=m_0 \phi^n e^{\beta \phi}$, where both power-law and exponential interactions contribute simultaneously. In this case, the effective potential exhibits characteristics of both forms: at small $\beta$, it closely resembles the power-law behaviour with minimal deviation, while at larger $\beta$, the exponential term dominates, causing a stronger rise in the potential and a noticeable steepening at higher $\phi$ values. The combined coupling thus introduces a tunable interaction strength, where $\beta$ governs the overall growth rate of the potential and $n$ provides additional modulation. This hybrid structure allows the model to interpolate smoothly between the exponential and power-law limits, leading to a more versatile inflationary behaviour that can accommodate a wide range of dynamical evolutions depending on the chosen coupling parameters.

The slow-roll parameters are evaluated as,

\begin{equation}
\epsilon' = \frac{1}{2}
\left(
\frac{
2V_0 \alpha \phi e^{-\alpha \phi^2}
+ \left( \beta + \frac{n}{\phi} \right)
\frac{1}{12}\left(\frac{m_\nu(\phi)}{T_\nu}\right)^2 
}{
V_0 \left( 1 - e^{-\alpha \phi^2} \right)
}
\right)^2.
\end{equation}

\begin{equation}
\eta =
\frac{
2V_0 \alpha e^{-\alpha \phi^2}(1 - 2\alpha \phi^2)
+ \left( \beta^2 + \frac{2\beta n}{\phi} + \frac{n(n - 1)}{\phi^2} \right)
\frac{1}{12}\left(\frac{m_\nu(\phi)}{T_\nu}\right)^2 
}{
V_0 \left( 1 - e^{-\alpha \phi^2} \right)
}.
\end{equation}

\noindent
The mixed coupling form generalizes the previous two cases, with the parameters $\beta$ and $n$ jointly determining the strength and nature of the scalar--neutrino interaction. While $\beta$ governs the exponential sensitivity of $m_\nu(\phi)$ to the scalar field, the power-law index $n$ introduces an additional field-dependent modulation. As a result, the effective potential’s slope and curvature, represented by the slow-roll parameters $\epsilon'$ and $\eta$, exhibit a richer dynamical behaviour, allowing for a broader range of inflationary and post-inflationary evolutions compared to the purely exponential or power-law couplings.

\begin{figure}[H]
\centering
\begin{minipage}[t]{0.45\textwidth}
\centering
\includegraphics[width=1.10\textwidth]{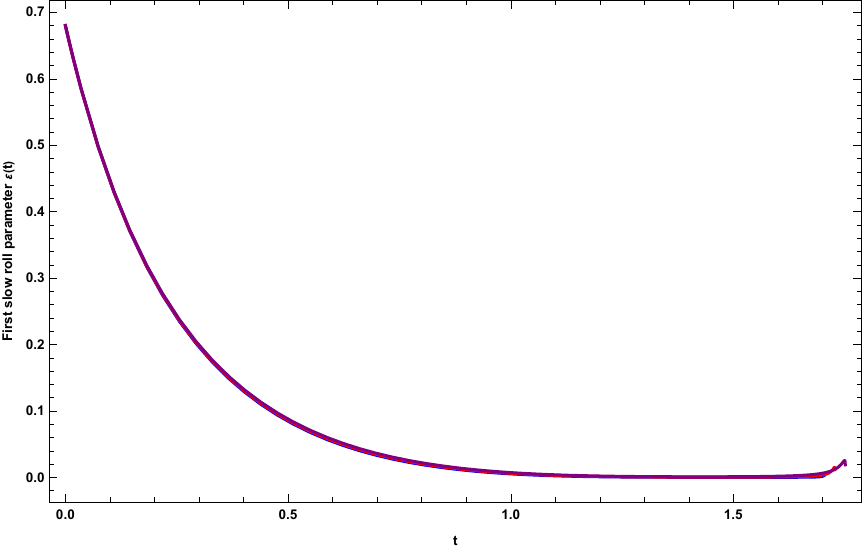}
\end{minipage}
\hfill
\begin{minipage}[b]{0.45\textwidth}
\centering
\includegraphics[width=1.10\textwidth]{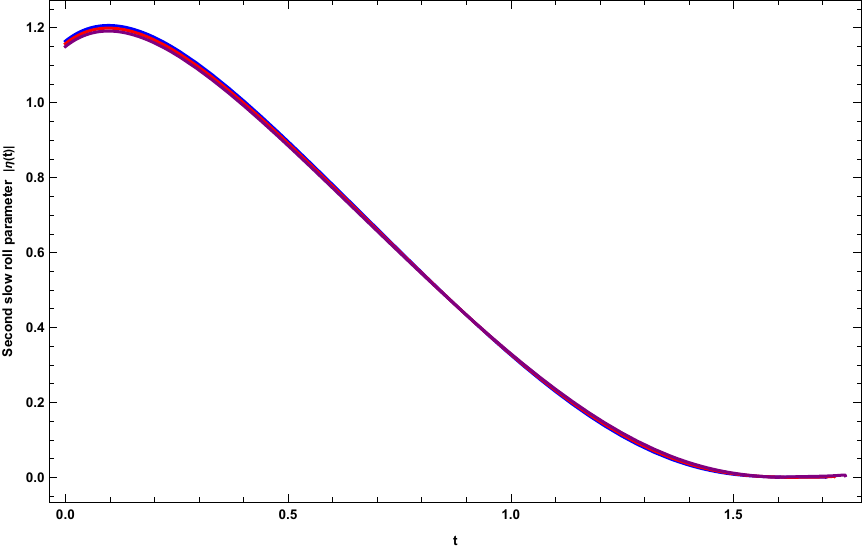}
\end{minipage}
\caption{Evolution of the first (left panel) and second (right panel) slow-roll parameters, $\epsilon'$ and $\eta$, respectively, for $V_0 = 0.18$ and $\alpha = 1$. The \textit{blue}, \textit{red}, and \textit{purple} curves correspond to the coupling parameters $\beta = 0.01$, $\beta = 0.03$, and $\beta = 0.05$, respectively, with $n = 0.01$.}\label{slowcomb}
\end{figure}

\noindent
Figure~\ref{slowcomb} presents the behaviour of $\epsilon'$ and $\eta$ for the mixed coupling case $m_\nu(\phi) = m_0 \phi^n e^{\beta \phi}$ with $n = 0.01$. This scenario combines the effects of both exponential and power-law couplings, leading to a richer and more flexible dynamical evolution. For small $\beta$, the behaviour closely resembles the power-law case, with $\epsilon'$ and $\eta$ remaining small for an extended period. As $\beta$ increases, the exponential term dominates, causing a rapid rise in both parameters and a quicker end to inflation. Therefore, the mixed coupling model provides a smooth transition between the power-law and exponential behaviours, with the parameters $\beta$ and $n$ jointly controlling the slope, curvature, and overall duration of the inflationary phase.

\section{conclusion}\label{con}

In this work, we explore  cosmological framework where neutrino masses arise dynamically through their coupling with a scalar field that also drives the inflationary expansion of the Universe. We begin with the Einstein--Hilbert action in a spatially flat FLRW spacetime, and derive the Friedmann and Raychaudhuri equations, in the presence of mass-varying neutrinos as a matter component. The modified continuity relations describe the energy exchange between the scalar field and the neutrino sector, illustrating how their interaction affects the overall dynamics. This coupling, introduced through the conformal function $B(\phi)$, leads to a nontrivial evolution of the energy densities and thereby alters the standard cosmic expansion history. When the neutrino mass depends explicitly on the scalar field, the model naturally gives rise to a class of mass-varying neutrino (MaVaN) scenarios, providing a unified picture in which neutrino mass generation and cosmological evolution are deeply interconnected.

We examined the kinetic description of neutrinos using the collisionless Boltzmann equation and derived expressions for their energy density and pressure, highlighting their dependence on the field-dependent mass $m_\nu(\phi)$. The resulting continuity equation shows that the neutrino energy density is not conserved independently, owing to a continuous energy exchange with the scalar field. This interaction introduces a coupling term proportional to $(\rho_\nu - 3p_\nu)$ in the scalar field dynamics, leading to the emergence of an effective potential $V_{\text{eff}}(\phi)$ that encapsulates the backreaction from the neutrino sector. For the relativistic regime, where $\rho_\nu - 3p_\nu \approx 0$, the contribution from mass-varying neutrinos is negligible, while in the nonrelativistic limit, this term effectively becomes proportional to $\rho_\nu$, significantly influencing the field evolution.

In our analysis, instead of treating these two limits separately, we considered a generalized formulation by explicitly evaluating the Fermi--Dirac integrals for the neutrino energy density and pressure, thus capturing the continuous transition between relativistic and nonrelativistic regimes. We then investigated the inflationary dynamics arising from the scalar--neutrino coupling by analyzing the evolution of slow-roll parameters within the effective potential $V_{\text{eff}}(\phi)$. During the inflationary epoch, corresponding to $\zeta = \frac{m_\nu(\phi)}{T_\nu} < 1$, we adopted a Gaussian-type scalar potential and explored three representative forms of the neutrino mass: exponential, power-law, and mixed couplings. For each case, we derived the slow-roll parameters $\epsilon'$ and $\eta$, and analyzed how different coupling choices affect the inflationary dynamics. The scalar--neutrino interaction introduces field-dependent corrections that modify both the slope and curvature of the potential. In the exponential coupling case, characterized by the parameter $\beta$, and in the power-law coupling case governed by the index $n$, we find that the framework successfully leads to an inflationary phase. Furthermore, the mixed coupling form, $m_\nu(\phi) = m_0 \phi^n e^{\beta \phi}$, provides a unified setting that can interpolate between these two limits, offering a broader parameter space for realizing inflation.
 
The coupling between the scalar field and the neutrino crucial for both the inflationary and late-time acceleration phases of the Universe. By appropriately tuning the coupling parameters $(\beta, n)$, it becomes possible to regulate the duration of inflation, the slow-roll behavior, and the subsequent transition to accelerated expansion. This unified framework thus establishes a compelling link between neutrino mass generation, scalar-field dynamics, and the large-scale evolution of the cosmos. Future extensions of this work may include a detailed confrontation of the model with cosmological observables, such as the scalar spectral index and tensor-to-scalar ratio, as well as exploring post-inflationary reheating and possible late-time acceleration within the same mass-varying neutrino paradigm.

In future work, we will perform a detailed comparison of our model with observational data.

\bibliographystyle{elsarticle-num}
\bibliography{mybib.bib}

\begin{thebibliography}{10}
\expandafter\ifx\csname url\endcsname\relax
  \def\url#1{\texttt{#1}}\fi
\expandafter\ifx\csname urlprefix\endcsname\relax\def\urlprefix{URL }\fi
\expandafter\ifx\csname href\endcsname\relax
  \def\href#1#2{#2} \def\path#1{#1}\fi

\bibitem{aghanim2020planck}
N.~Aghanim, et~al., Planck 2018 results. vi. cosmological parameters, Astron.
  Astrophys 641 (2020) A6.

\bibitem{DiValentino:2024xsv}
E.~Di~Valentino, S.~Gariazzo, O.~Mena, {Neutrinos in Cosmology} (2024).

\bibitem{Lesgourgues_2012}
J.~Lesgourgues, S.~Pastor, Neutrino mass from cosmology, Advances in High
  Energy Physics 2012 (2012) 1–34.

\bibitem{Hannestad_2010}
S.~Hannestad, Neutrino physics from precision cosmology, Progress in Particle
  and Nuclear Physics 65 (2010) 185–208.

\bibitem{desi2025a}
D.~Collaboration, Desi 2024 results: Bao measurements from the first three
  years (2025).

\bibitem{desi2025b}
D.~Collaboration, Desi 2025 results: Testing dynamical dark energy with dr2
  (2025).

\bibitem{choudhury2024updated}
S.~R. Choudhury, T.~Okumura, Updated cosmological constraints in extended
  parameter space with planck pr4, desi baryon acoustic oscillations, and
  supernovae: Dynamical dark energy, neutrino masses, lensing anomaly, and the
  hubble tension, The Astrophysical Journal Letters 976~(1) (2024) L11.

\bibitem{Chaudhary:2025uzr}
H.~Chaudhary, S.~Capozziello, S.~Praharaj, S.~K.~J. Pacif, G.~Mustafa, {Is the
  {\ensuremath{\Lambda}}CDM model in crisis?}, JHEAp 50 (2026) 100507.

\bibitem{paliathanasis2025dark}
A.~Paliathanasis, Dark energy within the generalized uncertainty principle in
  light of desi dr2, arXiv preprint arXiv:2503.20896 (2025).

\bibitem{kaur2025dynamics}
J.~Kaur, S.~Pathak, M.~Khlopov, M.~Krasnov, M.~Sharma, Dynamics of scalar
  fields from a generalized form of lagrangian, Nuclear Physics B (2025)
  117010.

\bibitem{bhandari2025distortion}
G.~Bhandari, S.~Pathak, M.~Sharma, A.~Wang, Distortion of quintessence dynamics
  by the generalized uncertainty principle, Annals of Physics 473 (2025)
  169895.

\bibitem{Dolgov_2002}
A.~Dolgov, Neutrinos in cosmology, Physics Reports 370~(4–5) (2002)
  333–535.

\bibitem{LESGOURGUES_2006}
J.~Lesgourgues, S.~PASTOR, Massive neutrinos and cosmology, Physics Reports
  429~(6) (2006) 307–379.

\bibitem{Weinberg:1962zza}
S.~Weinberg, {Universal Neutrino Degeneracy}, Phys. Rev. 128 (1962) 1457--1473.

\bibitem{lesgourgues2006}
J.~Lesgourgues, S.~Pastor, Neutrino cosmology and structure formation, Phys.
  Rept. 429 (2006) 307.

\bibitem{grohs2016}
E.~Grohs, et~al., Neff in the early universe, Phys. Rev. D 93 (2016) 083522.

\bibitem{PDG2024}
S.~Navas, et~al., Review of particle physics, Phys. Rev. D 110 (2024) 030001.

\bibitem{Giganti_2018}
C.~Giganti, S.~Lavignac, M.~Zito, Neutrino oscillations: The rise of the pmns
  paradigm, Progress in Particle and Nuclear Physics 98 (2018) 1–54.

\bibitem{Gonzalez_Garcia_2008}
M.~Gonzalez-Garcia, M.~Maltoni, Phenomenology with massive neutrinos, Physics
  Reports 460~(1–3) (2008) 1–129.

\bibitem{brookfield2006cosmology}
A.~W. Brookfield, C.~Van~de Bruck, D.~Mota, D.~Tocchini-Valentini, Cosmology of
  mass-varying neutrinos driven by quintessence: theory and observations, Phys.
  Rev. D 73~(8) (2006) 083515.

\bibitem{gellmann2013complexspinorsunifiedtheories}
M.~Gell-Mann, P.~Ramond, R.~Slansky, Complex spinors and unified theories
  (2013).

\bibitem{Yanagida:1980xy}
T.~Yanagida, {Horizontal Symmetry and Masses of Neutrinos}, Prog. Theor. Phys.
  64 (1980) 1103.

\bibitem{PhysRevLett.44.912}
R.~N. Mohapatra, G.~Senjanovi\ifmmode~\acute{c}\else \'{c}\fi{}, Neutrino mass
  and spontaneous parity nonconservation, Phys. Rev. Lett. 44 (1980) 912--915.

\bibitem{Fardon:2003eh}
R.~Fardon, A.~E. Nelson, N.~Weiner, {Dark energy from mass varying neutrinos},
  JCAP 10 (2004) 005.

\bibitem{Gu:2003er}
P.~Gu, X.~Wang, X.~Zhang, {Dark energy and neutrino mass limits from
  baryogenesis}, Phys. Rev. D 68 (2003) 087301.

\bibitem{Farrar:2003uw}
G.~R. Farrar, P.~J.~E. Peebles, {Interacting dark matter and dark energy},
  Astrophys. J. 604 (2004) 1--11.

\bibitem{amendola2000}
L.~Amendola, Coupled quintessence, Phys. Rev. D 62 (2000) 043511.

\bibitem{Starobinsky:1980te}
A.~A. Starobinsky, {A New Type of Isotropic Cosmological Models Without
  Singularity}, Phys. Lett. B 91 (1980) 99--102.

\bibitem{Kallosh_2013}
R.~Kallosh, A.~Linde, Universality class in conformal inflation, JCAP 2013~(07)
  (2013) 002–002.

\bibitem{Farrar_2004}
G.~R. Farrar, P.~J.~E. Peebles, Interacting dark matter and dark energy, The
  Astrophysical Journal 604~(1) (2004) 1–11.

\bibitem{wetterich1994cosmonmodelasymptoticallyvanishing}
C.~Wetterich, The cosmon model for an asymptotically vanishing time-dependent
  cosmological ``constant'' (1994).

\bibitem{Brookfield_2006}
A.~W. Brookfield, C.~van de Bruck, D.~F. Mota, D.~Tocchini-Valentini,
  Cosmology with massive neutrinos coupled to dark energy, Phys. Rev. lett.
  96~(6) (2006).

\bibitem{Liddle:2000cg}
A.~R. Liddle, D.~H. Lyth, Cosmological Inflation and Large-Scale Structure,
  Cambridge University Press, 2000.

\bibitem{Guth:1981}
A.~H. Guth, Inflationary universe: A possible solution to the horizon and
  flatness problems, Phys. Rev. D 23 (1981) 347--356.

\bibitem{Planck:2018jri}
P.~Collaboration, P.~A.~R. Ade, et~al., Planck 2018 results. x. constraints on
  inflation, Astron. Astrophys. 641 (2020) A10.

\bibitem{Sadjadi_2017}
H.~M. Sadjadi, V.~Anari, Mass varying neutrinos, symmetry breaking, and cosmic
  acceleration, Phys. Rev. D 95~(12) (2017).

\end{thebibliography}

\end{document}